# Shape and size-dependent surface plasmonic resonances of liquid metal alloy (EGaIn) nanoparticles

*Sina Jamalzadegan[1], Mohammadreza Zare [1], Micah J. Dickens[1], Florian Schenk [2], Alireza Velayati[1], Maksym Yarema[2], Michael D. Dickey[1]\*, and Qingshan Wei[1]\**

[1] Department of Chemical and Biomolecular Engineering, North Carolina State University, Raleigh, NC 27695, USA

[2] Department of Information Technology and Electrical Engineering, ETH Zürich, 8092 Zürich, Switzerland

\* Co-corresponding: qwei3@ncsu.edu and mddickey@ncsu.edu

**Abstract**

Liquid metals (LM) are emerging plasmonic nanomaterials with transformable surface plasmon resonances (SPR) due to their liquid-like deformability. This study delves into the plasmonic properties of LM nanoparticles, with a focus on EGaIn (eutectic gallium-indium)-based materials. Leveraging Finite-Difference Time-Domain (FDTD) simulations and experimental validations for spherical liquid metal nanoparticles plasmonic properties, we explored the localized SPR (LSPR) effects of EGaIn nanoparticles with various shapes, including nanospheres, dimers, nanorods, nanodisks, nanoellipses, nanocubes, and nanocuboids, in the broad range of ultraviolet (UV)-visible-near infrared (NIR) spectrum. While EGaIn is conventionally known as a UV-active metal alloy, this study reveals unique LSPR features of EGaIn (e.g., higher order resonances, polar and quadrupolar modes) in the broader visible and NIR wavelength ranges, providing a comprehensive map of LSPR properties for different shapes of EGaIn nanoparticles. These findings offer new insights into the dependence of the optical properties of EGaIn nanoparticles on their geometries for diverse applications, ranging from biosensing, nanoelectronics, to optomechanical systems.

**Keywords:** Liquid metal, EGaIn, surface plasmon resonance (SPR), FDTD, biosensor

**Table of contents**

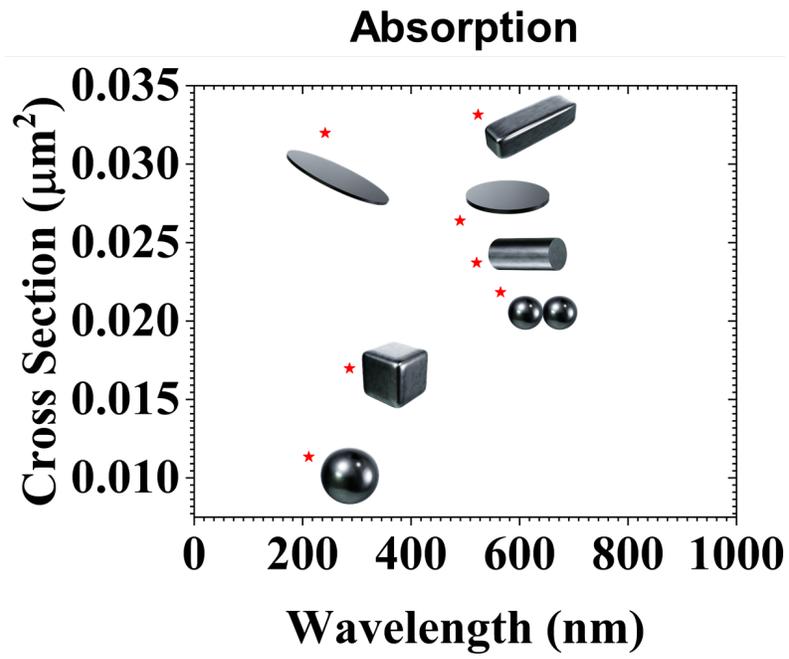

- **Introduction**

With the rapid advancement of nanofabrication methods, nanoplasmonics has experienced significant growth and found applications in a wide range of research areas. The surface plasmon resonance (SPR) phenomenon occurs between a conductive material and a dielectric medium. Under conditions of resonance, the incident light induces the formation of a collective charge density wave that propagates along this interface, known as a surface plasmon polarization (SPP)[1–3]. For metallic nanoparticles, the oscillation of electromagnetic waves is confined in close proximity to the nanoparticles themselves, resulting in an optical effect known as localized surface plasmon resonance (LSPR). The controlled manipulation of LSPR has given rise to a range of applications in areas such as biosensing[1,4,5], chemical detection[6], 3D printing[7], drug delivery[8], and the development of wearable devices[9]. For sensing applications, the sensitivity of a SPR sensor is based on the ability to detect small changes in the refractive index surrounding the plasmonic surface that cause shifts in the SPR[10]. The SPR shifts can be carefully measured and correspond to the presence and concentration of target analytes. The significance of SPR in biosensing lies in its ability to enable real-time and label-free detection of biomolecules, which makes it a potent diagnostic tool[11].

Gold and silver are the two most commonly studied plasmonic nanomaterials, which exhibit tunable LSPR properties spanning from visible (vis) to near infrared (NIR) wavelength range[12]. Alternative metals with LSPR features have also been reported. For instance, aluminum exhibits robust plasmon resonance across the ultraviolet (UV)-vis spectrum[13]. Platinum and palladium have been used for plasmonic optical gas sensing[14]. Copper could serve as plasmonic waveguides with exceptionally low losses, surpassing the performance of gold-based counterparts.[15] However, most conventional plasmonic nanomaterials are rigid, meaning mismatched mechanical properties when interfacing with biological surfaces. It also means fixed optical properties that cannot be flexibly modified based on the needs after synthesis or fabrication. On the other end, soft and stretchable plasmonic materials attract increased interests, because of the possibility of freely changing the shape, and thus, the SPR response of such materials. As such, soft and stretchable materials are finding various applications, ranging from soft robotics, e-skins, to stretchable electronics.

Gallium[16–20], indium[17,21], tin[17,21], magnesium[22], bismuth[23,24], and their alloy compositions[25,26] belong to the category of metals that exhibit plasmon resonance within the UV spectrum. Nanoparticles (NPs) fabricated from these materials may have different applications than gold and silver, which resonate in the visible range. For instance, plasmon-enhanced UV spectroscopy offers a valuable avenue to observe dynamic biochemical processes and characterize the chemical properties of these molecules in their native environment[27]. As such, UV-plasmonic NPs find utility in various fields, such as biomedicine[28], label-free DNA analysis[29], single molecule sensing[30], and cardiovascular imaging[31–33].

Liquid metals (LMs) consisting of gallium and its alloys have recently attracted attention as a UV-plasmonic material[16,18,21,34,35]. However, the UV region has not been the sole focus for LMs in recent years. Recent studies have highlighted the potential plasmonic properties of gallium alloy NPs in other wavelength regions, such as visible light and near-infrared (NIR)[36,37]. Despite this, a thorough investigation into how shape and size affect the surface plasmonic resonances of gallium alloy NPs is still lacking.

Gallium possesses a melting point of 29.8 °C, which can be lowered below room temperature by combining it with other metals like indium and tin. This alloying process results in eutectic compositions with remarkably low melting points. One prominent example of such eutectic composition is EGaIn (eutectic gallium-indium), consisting of 75 wt% gallium and 25 wt% indium, with a melting point of 15.7 °C. Furthermore, both gallium and EGaIn exhibit low toxicity, exhibit negligible vapor pressure at room temperature, have a viscosity approximately twice that of water, and possess excellent electrical and thermal conductivity[38–44]. Due to their low viscosity, EGaIn can be readily fragmented into colloidal droplets through various techniques[45]. Recent reviews have extensively examined the utilization of these droplets, including their applications in sensors, microfluidic systems, robotics, electronic circuits, catalysis, energy harvesting, and even biomedical applications like drug delivery[45–47]. Currently, the prevalent methods for generating LMNPs (liquid metal nanoparticles) are primarily based on employing microfluidic, sonication, and shearing processes[46].

Gallium exhibits a Drude-like dielectric permittivity profile that spans from the UV range [48] all the way into the NIR region[25,34]. Moreover, within nanoscale constraints, the oscillations of charge carriers can be readily affected by slight alterations in size, geometry, material composition, and variations in the surrounding dielectric environment in which the nanoparticles are embedded. Recent computational studies examined the LSPR sensitivity of rod-shaped core-shell GaAg, GaAl, GaHg, and Ga (85.8%) In (14.2%) (which is not EGaIn) NPs. These studies showed a red shift of the extinction spectrum peaks of GaAg from the UV to the visible light range by varying the aspect ratio and shell thickness. In contrast, Ga NPs alone exhibited dominant peaks in scattering and absorption power cross-section spectra in the UV wavelength range. Additionally, Ga (85.8%) In (14.2%) NPs did not show a substantial LSPR effect compared to GaAg[49–51]. Although these computational studies focused on gallium alloys, they still lack computational findings on EGaIn surface plasmon resonance features. In the limited examples in the past, the UV plasmonic features of 100 nm diameter EGaIn NPs have been experimentally and computationally confirmed[52]. Nevertheless, a thorough study on the correlation of the LM shape and size with their LSPR properties has not been presented.

Gallium and its liquid alloys are appealing due to the possibility of shape-transformation and constructing reconfigurable plasmonic devices. Thus, it is critical to substantiate theoretical projections related to gallium nanoparticles concerning their dimensions, geometry, response to the surrounding dielectric environment, and their integration with other plasmonic materials, leading to the controllable optical and plasmonic characteristics. Furthermore, with improvements in the fabrication[53,54] and synthesis of liquid metal nanodroplets with distinct shapes, it is imperative to explore how the various nanoparticle geometries influence the manifestation of their unique properties. We focus on EGaIn as a representative LM to illustrate how "liquid plasmonics" can be rationally programmed by controlling the size and shape. By combining its robust SPR properties, geometry-tailored plasmonic effects, and outstanding non-optical properties, EGaIn nanoplasmonics offers exciting opportunities for applications in photoacoustics, nanoelectronics, nanoscale electrochemistry[55], and optomechanical systems[56].

In this study, we applied Finite-Difference Time-Domain (FDTD) numerical simulation technique[57] to investigate the geometry-LSPR relationship of EGaIn-based LM nanomaterials. We investigated the scattering and absorption cross-section spectra of EGaIn NPs with different shapes and sizes, as well as their coupling with gold and silver NPs, to identify characteristic plasmon-resonant peaks in the wide UV-vis-NIR range. We also confirmed our simulation results for the plasmonic features of spherical EGaIn and Ga nanoparticles through experimental fabrication and optical analysis of these two types of nanoparticles. This validation is in addition to previous literature and other theoretical simulations. While dominant plasmon-resonant activities of EGaIn NPs were mostly observed in the UV wavelengths as expected, new plasmonic activities in the visible wavelength range were also revealed by controlling EGaIn shapes that were not experimentally reported before. The results can be used to guide the future synthesis of EGaIn NPs with desired optical properties that are useful for various optical biosensing applications.

- **Methods**

FDTD simulation. The FDTD numerical simulation was run by the Lumerical Ansys software (2024 R2.1). The permittivity of EGaIn in the 400–1000 nm range reported in the literature[25] was extrapolated[58] to the UV range of 100–400 nm using the Drude equations (**Equations 1 and 2**). For FDTD simulations spanning the ultraviolet (UV), visible, and near-infrared (NIR) spectral ranges, the Drude free-electron model for EGaIn[59] proves to be a simplified yet efficient approach for modeling the optical response of metals dominated by free-electron (intraband) dynamics[60,61].

$$\varepsilon_1(E) = 1 - \frac{E_p^2}{(E^2 + \gamma^2)} \quad (1)$$

$$\varepsilon_2(E) = \frac{E_p^2 \gamma}{E(E^2 + \gamma^2)} \quad (2)$$

where $\varepsilon_1$ and $\varepsilon_2$ represent the real and imaginary part of the permittivity, respectively. $E_p$ and $\gamma$ represent the plasma frequency and broadening parameters. These two parameters were calculated by fitting the experimental refractive index data[25] using the Drude free-electron model. **Figure S1** shows the extrapolated and experimental real and imaginary parts of the permittivity of EGaIn in the 100–1000 nm wavelength range. The simulation time was set to 1000 fs at 300 K (to reflect standard ambient laboratory conditions, which are typically around room temperature), and the refractive index of the surrounding medium was set to 1, which corresponds to air. For DMF and ethanol, we chose refractive indexes of 1.43 and 1.478, respectively. The model ignores any surface oxides or adsorbate molecules. The mesh size was set uniformly at 0.5 nm. A s-polarized (90° polarization angle) plane source with 100–1000 nm wavelength was used to study the absorption and scattering power cross-section responses. **Figure S2** schematically shows the FDTD simulation setup with the incident light source and polarization angle.

**Mie Theory.** To validate our FDTD findings, we also simulated the absorption and scattering cross-section spectra with Mie theory[62] and compared with FDTD results. The formulas presented in **Equations 3 and 4** represent the scattering and absorption cross-sections ($\sigma$) derived from the Mie theory, where a is the radius of a spherical nanoparticle and $\alpha = \frac{2\pi a}{\lambda}$. Also, $a_n$ and $b_n$ are the Mie theory coefficients, which represent the magnetic and electric poles of order n, respectively.

$$\sigma_{absorption} = \frac{\pi a^2}{\alpha^2} \left| \sum_{n=1}^{\infty} (2n+1)(-1)^n (a_n - b_n) \right|^2 \tag{3}$$

$$\sigma_{scattering} = \frac{2\pi a^2}{\alpha^2} \sum_{n=1}^{\infty} (2n+1)(|a_n|^2 + |b_n|^2) \tag{4}$$

- **EXPERIMENTAL SECTION**

**Synthesis of spherical EGaIn and Ga nanoparticles.** The spherical EGaIn nanoparticles were synthesized by following the protocol described previously.[51] Briefly, 200 mg of EGaIn was added to 10 mL of solvent (either ethanol or dimethylformamide (DMF)) in a 20 ml vial. The vial was then placed in an ice bath for temperature control. The solution was sonicated with a 1/8-inch microprobe tip at 80 amplitude for 20 minutes on a 2-second on and 2-second off cycle. Large particles were removed by slow centrifugation (500-1000 rpm for 5 min) and the supernatant that contained small particles were collected for future experiments.

Ga nanoparticles were synthesized following the protocols described in a previous publication.[68] For 42-nm Ga nanoparticles, 21 mL 1- octadecene were loaded in a 100 mL 3-neck flask, equipped with a Liebig condenser and dried under dynamic vacuum at 110 °C for 1 h. The flask was filled with nitrogen and heated to 280 °C, followed by injection of a solution of 37.5 mg $Ga_2(DMA)_6$ in dried dioctylamine (1.695 mL) and 16.3 mL 1- octadecene. The temperature dropped to ca. 205 °C and quickly recovered to 230 °C. After 20 min, the flask was cooled to temperature. For purification, 20 mL chloroform, 0.2 mL oleic acid and 40 mL ethanol were added, followed by centrifugation. The particles were redispersed in chloroform and the purification steps were repeated twice. For 27-nm Ga nanoparticles, the same synthesis and purification steps were used, except the injected solution contained 75 mg $Ga_2(DMA)_6$, dried didodecylamine (3.9 g) and 12.6 mL 1- octadecene and the growth time was 1 min.

**LM nanoparticle characterization.** For UV-vis absorption measurement, 300-500 μL of EGaIn nanoparticle solutions in ethanol or DMF were loaded to a standard quartz cuvette. Absorption spectra were obtained on a Thermo Scientific Evolution 201 UV-vis spectrometer with a 1 nm scanning resolution. For DLS measurements on the Malvern Zetasizer nano, 500 μL of EGaIn nanoparticle solution were added to 5 mL of DI water. Given that the particle surface is comprised of gallium, a refractive index of 1.950 with adsorption of 0.01 was chosen per the Zetasizer software. Transmission electron microscopy (TEM) was carried out on JEM-1400 JEOL microscope operated at 100 kV by dropping diluted nanoparticle dispersions on a carbon grid. Particle size distributions were obtained by measuring > 100 particles using the ImageJ software and fitting a Gaussian size distribution.

▪ **Results and discussion**

**SPR effects of 0D EGaIn nanoparticles**

*EGaIn Nanospheres.* A nanosphere is a zero-dimensional (0D) nanoparticle characterized by its spherical shape. Nanospheres have uniform diameters that typically range from a few nanometers to hundreds of nanometers. Their properties and behaviors often differ from those of bulk materials due to quantum effects and size-related phenomena. Studying LM nanospheresis sensible because liquids naturally like to assume spherical shapes due to surface tension.

As an initial step, we examined the scattering and absorption cross-sections of individual spherical EGaIn nanoparticles with varying diameters ranging from 20 to 200 nm (**Figure 1**). Power cross-section is the probability of light energy being absorbed or scattered by the objects, which is the key parameter for assessing how effectively a material interacts with light, quantifying its energy absorption and scattering capabilities across wavelengths. Our study utilizes absorption and scattering power cross-sections across UV to near-IR wavelengthsfor characterizing various shapes and sizes of EGaIn nanoparticles (**Figure 1a**). Using Lumerical Ansys software, we extracted the computed electric field data for a 200-nm single spherical EGaIn particle at a 253 nm wavelength (major scattering peak) and plotted the electric field heat map using Python packages **(Figure 1b)**. For the smaller sizes of EGaIn nanoparticles (20-100 nm), the absorption cross-section spectra exhibit SPR effects mainly within the UV range (around 200 nm wavelength) (**Figure 1c**). The amplitude of the dominant LSPR in the absorption spectra increases with the increase of LM NP size, a phenomenon that was corroborated by a recent simulation study[52]. In addition, the absorption spectrum of a 100 nm spherical EGaIn nanoparticle displays dominant peaks within the UV range (**Figure 1c**), consistent with prior experiments and simulation model [52]. Furthermore, the FDTD simulation results indicate that as the NP size increases, a secondary

LSPR peak becomes more and more evident within the visible light spectrum (400-700 nm) and the near-infrared (IR) range (700-1000 nm) of the absorption spectra (**Figure 1c**). The visible SPR response of spherical EGaIn nanoparticles is even more obvious in the scattering spectra (**Figure 1d**). This observation indicates EGaIn NPs could be excellent scattering contrast agents in the visible wavelength range.

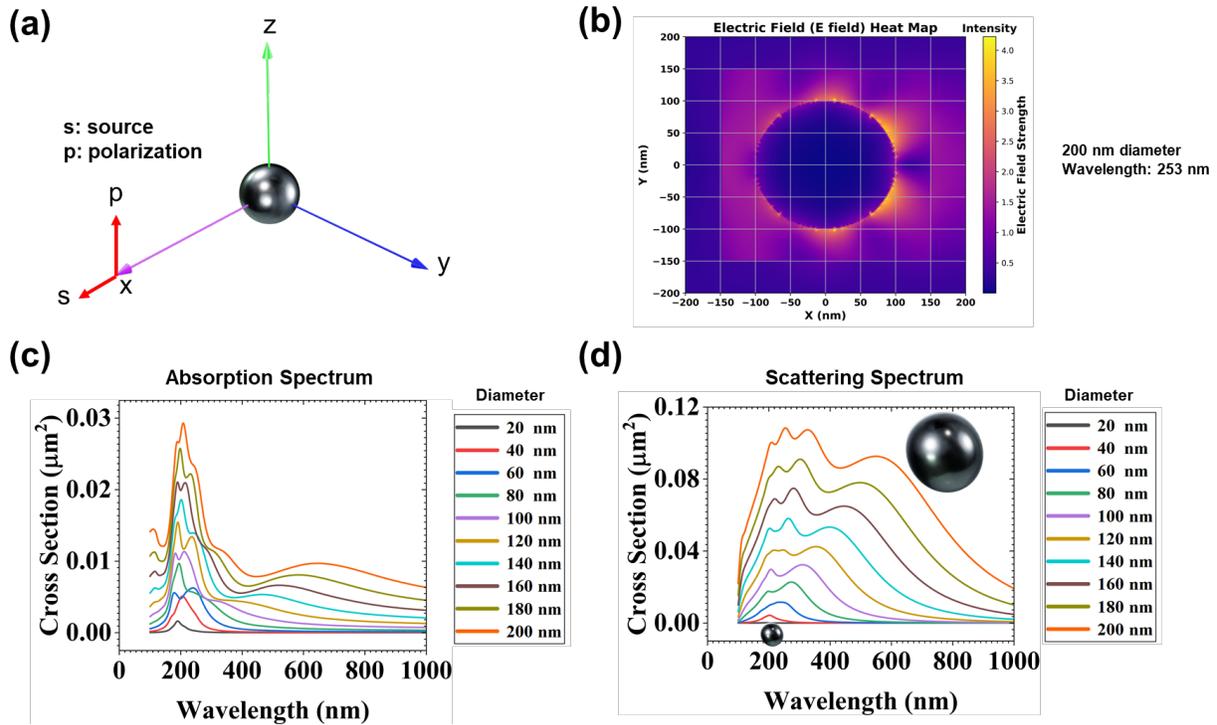

**Figure 1. Plasmonic resonance behavior of a single spherical EGaIn nanoparticle.** (a) 3D schematic of a single spherical EGaIn NP excited by a plane-wave light source (s) with 90º polarized angle (p). (b) Electrical field (E field) heat map for a 200-nm diameter spherical EGaIn nanoparticle excited at 253-nm wavelength (corresponding to its major scattering peak). The plane-wave light source propagates in x-direction from left to right. (c) Absorption and (d) scattering cross-section spectra of single spherical EGaIn nanoparticles with various diameters (20-200 nm).

The cross-section spectra obtained using Mie theory exhibit a close correspondence with the FDTD results (**Figure S3**). LMs are known to have an ultrathin (~3 nm) oxide layer, which is naturally formed under ambient conditions. We also studied the effects of the oxide layer on the LSPR properties of EGaIn NPs. We ran a simulation of a 200-nm spherical EGaIn NP with a 6-nm oxide layer. The FDTD simulation results showed that the thin oxide layer did not impact the plasmonic features of the EGaIn NP, as no obvious changes were observed in the absorption and scattering spectra compared to the pure EGaIn NPs without the oxide layer (**Figure S4**). Furthermore, we explored the SPR effects of single gallium (Ga) spherical NPs (**Figure S5**). Similar to the single EGaIn NPs, we observed that the LSPR of the single Ga NPs shows a prominent peak in the UV range in both the absorption and scattering cross-section spectra. Like EGaIn NPs, the scattering cross-section spectra of Ga NPs also have a secondary peak within the visible light range and its position redshifts as the particle size increases. Despite the presence of indium in the eutectic alloy, the bulk dielectric properties of EGaIn remain dominated by gallium due to their similar Drude-like behavior and free-electron responses in the UV-visible range. Although indium enriches the EGaIn surface, as reported in high-vacuum studies, ellipsometry measurements under oxide-free, electrochemically reduced conditions show that this enrichment does not substantially impact the bulk optical constants[25]. As such, the LSPR features of EGaIn nanospheres closely resemble those of pure Ga, with only a modest red-shift attributable to slight differences in plasma frequency and damping. It is worth mentioning that compared to EGaIn, gallium has a melting point slightly above room temperature. Yet, it is typically found as a liquid at room temperature due to its ability to supercool[63]. The supercooling is thought to arise, in part, because of the oxide layer and also the effect of using particles to limit heterogeneous nucleation.

***EGaIn Nanocubes***. Nano-sized cubic-shaped NPs exhibit unique properties owing to their well-defined structures and sharp corners. These properties include a high surface area-to-volume ratio, making them excellent candidates for catalytic applications, as well as efficient drug delivery carriers due to their enhanced surface reactivity. Their distinct geometry leads to plasmonic enhancement (e.g., hot spots), enabling applications in sensing and imaging. Additionally, the crystalline nature of cubic nanoparticles ensures remarkable stability and controlled surface chemistry, facilitating precise tuning of their electronic, optical, and magnetic characteristics. These properties make nano-sized cubic-shaped nanoparticles invaluable in a wide range of fields. While fabricating EGaIn NPs with cubic shapes poses a significant challenge due to the high surface tension of EGaIn, our study focused on investigating the SPR effects of these NPs for potential future research endeavors in this field.

**Figure 2** displays the absorption and scattering power cross-section spectra for cubic EGaIn NPs ranging from 50 to 400 nm in cubic size. Using Lumerical Ansys software, we extracted the electric field data for a 400-nm single EGaIn nanocube at 215 nm and 970 nm wavelengths (major scattering peaks), respectively, and plotted the electric field heat maps of both using Python packages **(Figure 2b)**. The 2D field maps showed enhanced LSPR effects (or hot spots) either at the edges or corners of nanocubes, depending on the excitation wavelengths. The absorption power cross-section spectrum reveals the presence of two plasmon resonance response regions: one in the UV range (mode I, around 200 nm), characteristic of EGaIn, and the other one in the vis-NIR range (mode II), which is observed for nanocubes larger than 100 nm size (**Figure 2c**). The model I peak observed for EGaIn nanocubes in the UV range is consistent with other metal nanoparticles, such as silver nanocubes, which exhibit a main SPR peak in the visible light range[64,65]. The model I peak also consists of several small peaks. Notably, as the size of the cubic-shaped EGaIn nanoparticles increases, resonant mode I of nanocubes remains at around 200 nm, while mode II peaks exhibit a redshift all the way into the NIR region with increased peak intensity (**Figure 2c**). This dual-peak mode plasmonic behavior was previously investigated in silver nanocubes and attributed to the variation in surface charge localization across distinct geometric features, namely the shorter wavelength quadrupolar SPR localized to the corners of nanocubes and longer wavelength dipolar SPR localized along the sides of nanocubes [66]. The scattering power cross-section spectra also demonstrated two major peaks in the UV and vis-NIR regions, respectively. However, the latter has a much higher amplitude than the UV resonant mode, indicating EGaIn nanocubes are better scatters in the vis-NIR region compared with UV wavelengths (**Figure 2d**).

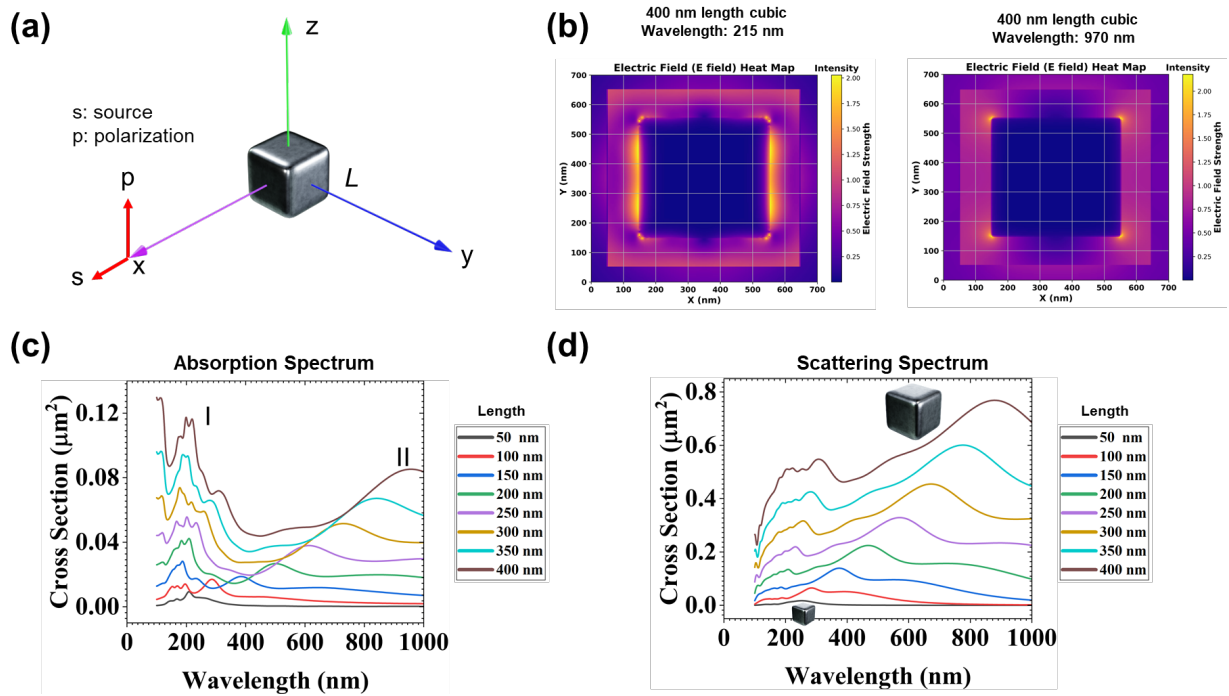

**Figure 2. Plasmonic resonance behavior of single cubic EGaIn nanoparticle. (a)** 3D schematic of single cubic EGaIn nanoparticle excited by a plane-wave light source (s) in x-direction with a 90 degree polarization (p) angle. **(b)** Electrical field (E field) heat map for a 400-nm length cubic

EGaIn nanoparticle excited at 215 nm and 970 nm wavelength (major absorption peaks), respectively. The plane-wave light source propagates in x-direction from left to right. (c) Absorption and (d) scattering cross-section spectrums of EGaIn nanocubes with various lengths (50-400 nm).

**SPR effects of 1D EGaIn nanoparticles**

*EGaIn Nanorods.* Here, a cylinder is chosen as a simple model for FDTD simulation of 1D LM materials. Nanorods may refer to any low aspect ratio (AR) 1D nanomaterials with many other noncircular cross-section shapes, such as square or pentagon. When the AR exceeds 10-20, these 1D nanomaterials are also generally referred to as nanowires. EGaIn nanorods, measuring 210 nm in diameter and 850 nm in length, were previously synthesized via an ultrasound-assisted physical dispersion technique[67]. In a different method, EGaIn droplets anchored to a flat substrate were pulled perpendicular to the substrate surface at room temperature, resulting in the formation of an hourglass shaped EGaIn structure. Subsequent pulling leads to the breakdown of this bridge, ultimately resulting in the formation of EGaIn NWs on the surface at the point of rupture[68]. Thus, different from the nanocubes, 1D LM materials have already been realistically prepared.

We investigated the surface plasmon resonance of single EGaIn nanorods by altering both the diameter and length of individual EGaIn nanorods. **Figure 3a** illustrates the schematic representation of a single EGaIn nanorod, with a constant diameter (20 nm) while varying its length (20-1000 nm). **Figures 3a, d,** and **g** display the exposure of a single EGaIn nanorod to a polarized plane-wave source in all three directions. The results of the absorption (**Figures 3e and 3h**) and scattering (**Figures 3f and 3i**) cross-sections for both the y and z plane-wave sources revealed two prominent peaks in the UV range. Additionally, as the length of the nanorods increased, the LSPR peaks of both scattering and absorption cross-section spectra also increased. Notably, when switching to an x-direction plane-wave source, we observed a dominant LSPR peak shifted into the visible light range for both absorption (**Figure 3b**) and scattering (**Figure 3c**) cross-section spectra, while a weak LSPR peak remained in the short wavelength range (<400 nm). We attributed the short-wavelength LSPR features observed under polarization excitation (Fig. 3b,c) to higher-order longitudinal modes rather than transverse modes, which are typically excited under perpendicular polarization (as in Fig. 3e,f,h,i). Additionally, for EGaIn nanorods with lengths approaching or exceeding the excitation wavelength (~1000 nm), retardation effects become significant. The phase of the incident electromagnetic field is no longer uniform across the structure, which leads to destructive interference between different segments of the rod and diminishes the efficiency of coherent electron oscillation along the longitudinal axis. This disrupts the collective resonance and suppresses the dipolar LSPR signature (**Figure 3b**).

By maintaining a constant length (1000 nm) while varying the diameter of these nanorods, we observed SPR effects mainly confined in the UV range (**Figure S6**). As anticipated, both the scattering and absorption cross-sections increased with an increase in the radius of these nanorods (**Figure S6**).

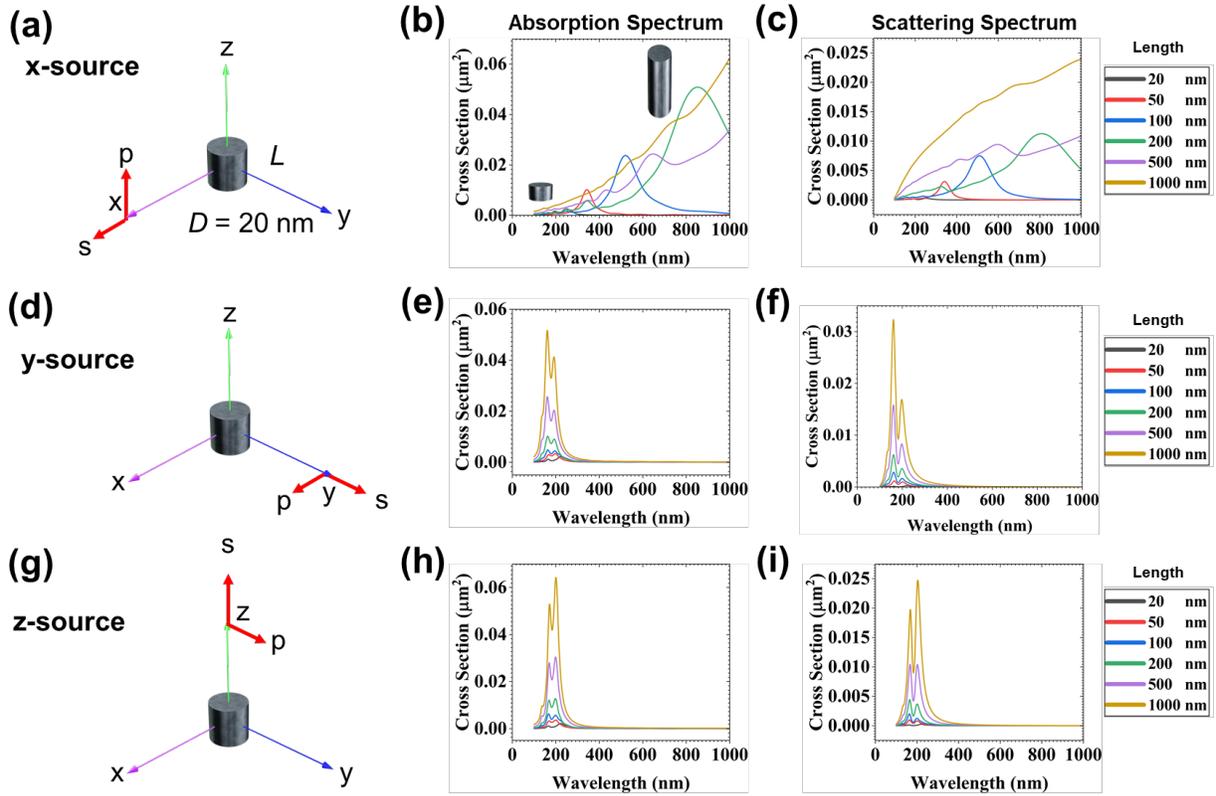

**Figure 3. Plasmonic resonance behavior of single EGaIn nanorod.** (a), (d), and (g) 3D schematic of single EGaIn nanorod excited by a plane-wave light source (s) in x-, y-, and z-directions with 90 degree polarization (p), respectively. (b), (e), and (h) Absorption and (c), (f), and (i) scattering cross-section spectra for various lengths with a constant diameter of 20 nm.

*EGaIn Nanocuboids.* Nanocuboid is another type of 1D EGaIn, which has a square cross-section and tunable length. Such a structure has not been fabricated yet for EGaIn. To investigate the LSPR effects of EGaIn nanocuboids, we simulated the absorption and scattering power cross-section spectra using plane-wave sources in all three directions for single EGaIn nanocuboids with a 100 x 100 nm cross-section and various lengths from 5 to 500 nm (**Figure 4**). For the absorption (**Figure 4b**) and scattering (**Figure 4c**) spectra of the single EGaIn nanocuboid excited by a plane-wave light source in the x-direction (**Figure 4a**), a multi-mode LSPR response was observed, with the major LSPR peak located in the vis-NIR wavelength range. Similar to the LM nanorods, this dominant peak is attributed to the longitudinal resonance of the nanocuboids.

For y and z-direction excitation (Figures 4d and 4g), the major peaks of the absorption (**Figures 4e** and **4h**) and scattering spectra (**Figures 4f** and **4i**) showed a blue-shift into the UV range as the cuboid length increases. For the absorption (**Figure 4e**) spectra of the single EGaIn nanocuboid excited by a plane-wave light source in the y-direction (**Figure 4d**), the UV LSPR peak also spitted into two for lengths larger than 20 nm. We attributed the multi-mode LSPR responses of nanocuboids to the strong edge effects similar to the previous case of nanocubes, where the sharp corners introduce shorter wavelength quadrupolar resonance and the long sides contribute to the longer wavelength dipolar response.

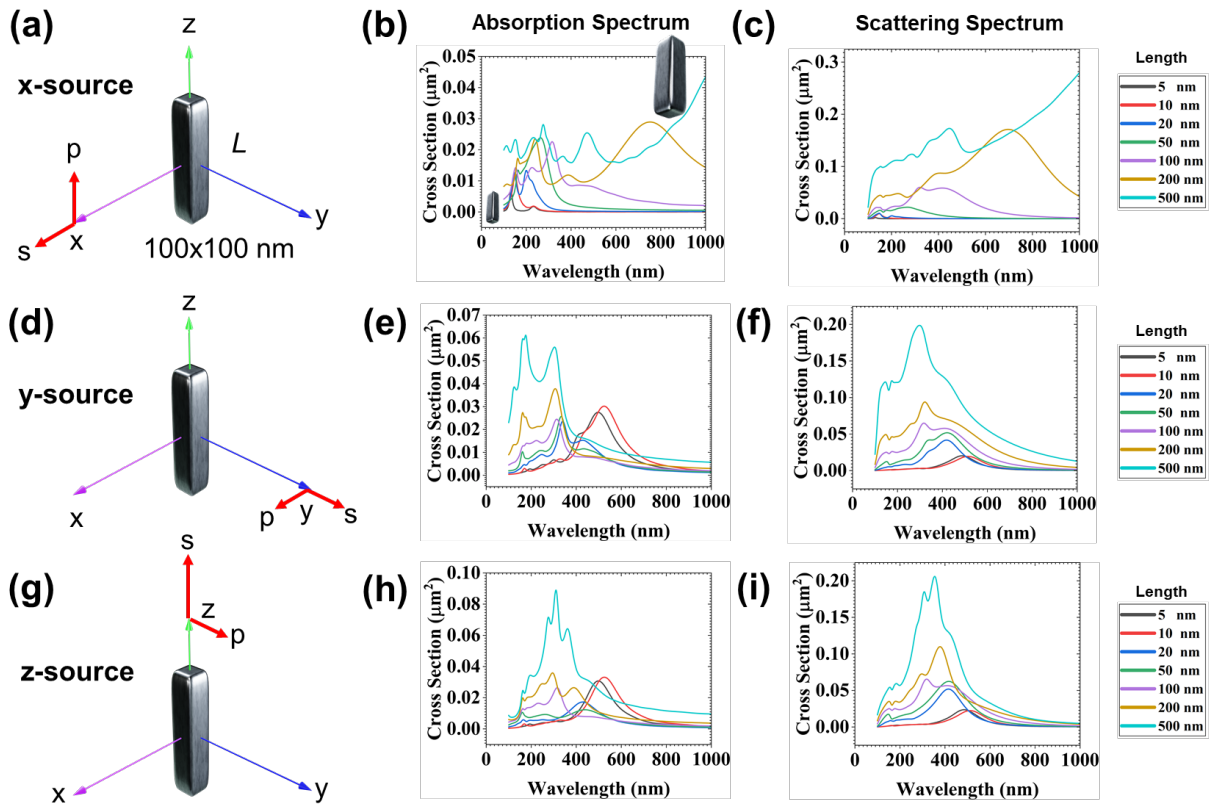

**Figure 4. Plasmonic resonance behavior of single EGaIn nanocuboid.** (a), (d), and (g) 3D schematic of a single EGaIn nanocuboid excited by a plane-wave light source (s) in x-, y-, and z-direction with 90 degree polarization (p), respectively. (b), (e), and (h) Absorption and (c), (f), and (i) scattering cross-section spectra for various lengths with constant cross-section area of 100x100 nm.

**SPR effects of 2D EGaIn nanoparticles**

*EGaIn Nanodisks.* 2D EGaIn nanostructures include several shapes, such as nanodisks and nanoellipses. Nanodisks are disk-like structures with two flat surfaces. **Figure 5** illustrates the absorption and scattering cross-sections of a single EGaIn nanodisk (5 nm thick) under the excitation of three distinct plane-wave source directions. When utilizing an x-direction plane-wave source, both the absorption (**Figure 5b**) and scattering (**Figure 5c**) cross-section spectra exhibited a single SPR peak of EGaIn nanodisks in the UV range. As the diameter of EGaIn nanodisk increased, the peak experienced a slight redshift, accompanied by an dramatic increase in the power of cross-sections. When employing a y-direction plane-wave source, the previous UV LSPR peaks disappeared (**Figures 5e and 5f**). Instead, for larger nanodisks, such as 100 nm and 200 nm diameter nanodisks, broad SPR peaks appeared in the visible and NIR wavelength ranges for both absorption (**Figure 5e**) and scattering (**Figure 5f**) spectra. The intensity of such peaks increased with the increase of nanodisk diameter. In a similar manner, the z-direction plane-wave source produced a prominent SPR peak in the visible and NIR regions for both absorption (**Figure 5h**) and scattering (**Figure 5i**) spectra. These results suggested that EGaIn nanodisks could be very active NIR-responsive plasmonic materials under right excitation.

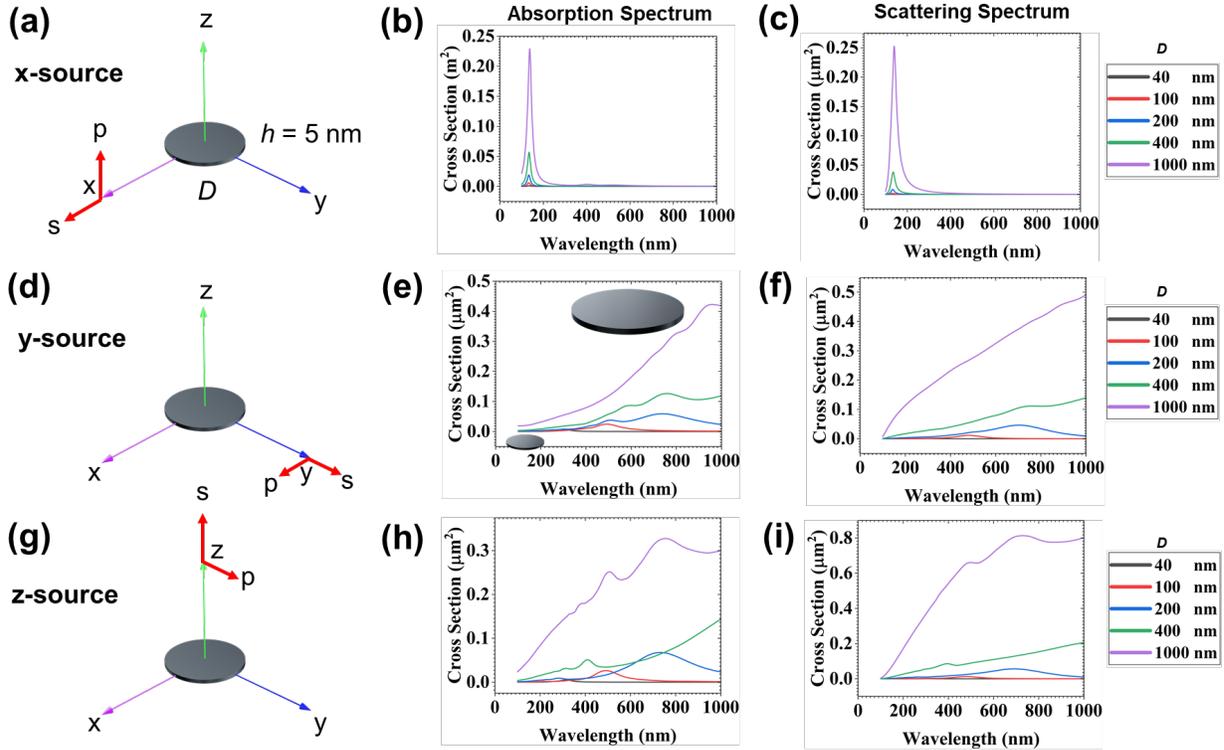

**Figure 5. Plasmonic resonance behavior of single EGaIn nanodisk.** (a), (d), and (g) 3D schematic of single EGaIn nanodisk excited by a plane-wave light source (s) in x-, y-, and z-directions with 90 degree polarization (p), respectively. (b), (e), and (h) Absorption and (c), (f), and (i) scattering cross-section spectrums for various diameters with fixed 5 nm thickness.

*EGaIn Nanoellipses*. Nanoellipses are 2D structures with two asymmetric axes (a and b). Previously, laser ablation of gallium metal was utilized to synthesize ellipse-like GaOOH particles in an aqueous solution. The successful formation of well-defined GaOOH structures was controlled by the addition of cationic CTAB surfactant[70]. Fabrication of EGaIn nanoellipses has not been demonstrated before. However, it is essential to study the LSPR of EGaIn nanoellipsoids to explore their unknown optical properties and corresponding applications in various fields. We investigated the LSPR effects of single EGaIn nanoellipses by fixing the thickness (50 nm) and the length of minor axis (b=100 nm, **Figure S7**). **Figure S7** presents the absorption and scattering cross-section spectra of this NP shape. By applying a plane-wave source in the x-direction (**Figure S7a**), we observed LSPR effects of EGaIn nanoellipses mainly in the UV range for ellipsoids with varied major axis, ranging from 100 to 1000 nm, while keeping the minor axis constant at 100 nm (**Figure S7b and S7c**). While the dominant peaks are located at ~300 nm, there is a notable shoulder peak at ~150 nm in the deep UV range. Additionally, the power cross-section for both absorption and scattering spectra increased with the increase of the ellipse's major axis. When applying a plane-wave source in the z-direction (**Figure S7g**), we observed a similar plasmon resonance behavior in the UV range (**Figure S7h and S7i**). Interestingly, when applying a plane-wave source in the y-direction (**Figure S7d**), we observed multimodal LSPR responses of single EGaIn nanoellipses with radius sizes of long axis in the broad spectra of visible and NIR wavelengths (**Figure S7e and S7f**). The longer the major axis, the more LSPR modes the nanoellipses have. We attributed the increasing number of resonant modes to the transition of oval shapes of nanoellipses into 1D geometries when the ratio of major to minor axis increases.

**Plasmonic Coupling of EGaIn NPs**

*EGaIn-EGaIn coupling.* In addition to particle size and shape, the distance of neighboring plasmonic NPs will also affect its own LSPR property due to SPR overlapping, an effect also known as plasmonic coupling. The plasmonic coupling effect of noble metals such as gold[71] and silver[72] has been well studied previously by employing spherical nanoparticle dimers as models. In a similar vein, here we investigated the SPR coupling effects of spherical dimer NPs made of EGaIn. In the FDTD simulation, we constructed EGaIn dimers with different particle sizes (**Figure 6**). Using Lumerical Ansys software, we extracted the computed electric field data for a 100-nm dimer spherical EGaIn particles excited by a 210 nm wavelength plane-wave in all directions with 90 degree polarization and plotted the electric field heat map **(Figure 6b, 6f, and 6j)**. The results showed that by changing the orientation of the excitation source, it is possible to control the hot spot locations, either in between or at the outside edge of the dimers. The whole spectra results indicate that when employing a plane-wave source oriented along the x (**Figures 6c, and 6d**) and z (**Figures 6g and 6h**) directions, mild plasmonic coupling occurred for the dimers with a new SPR peak appearing at around 200-300 nm. Moreover, these peaks exhibited a redshift as the dimer size increased. Notably, when the plane-wave source was oriented along the y direction, the coupling effect was more significant, with a dominant coupling peak emerging within the visible light and NIR spectrum (**Figures 6g and 6h**). Again, a larger dimer size correlates with redshifted and enhanced resonance-coupled peaks.

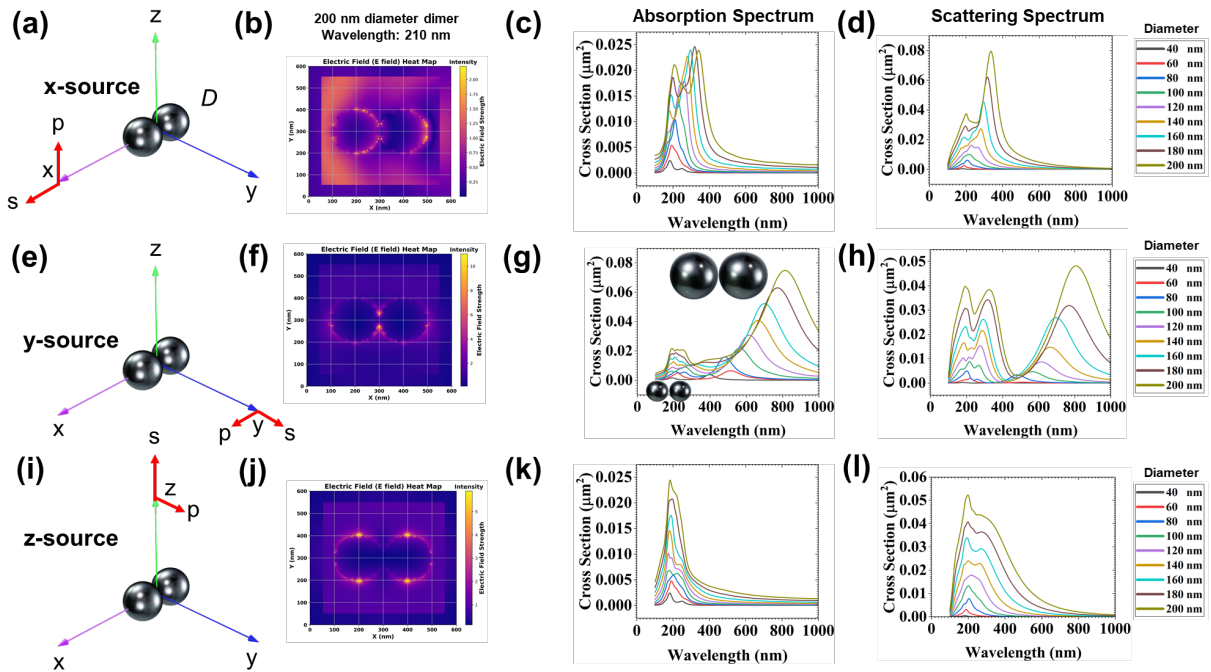

**Figure 6. Plasmonic resonance behavior of spherical EGaIn dimers.** (a), (e), and (i) 3D schematic of spherical EGaIn dimers excited by a plane-wave light source (s) in x-, y-, and z-direction with 90 degree polarization (p), respectively. (b), (f), and (j) Electrical field (E field) heat map for 100 nm diameter spherical EGaIn nanoparticle dimer at 210 nm wavelength excited by a plane-wave light source (s) in x-,y-, and z-direction with 90 degree polarization (p), respectively. (c), (g), and (k) Absorption and (d), (h), and (l) scattering cross-section spectra for EGaIn dimers with various individual particle sizes.

*EGaIn-Au and EGaIn-Ag coupling*. Additionally, we investigated the LSPR coupling behavior of EGaIn-noble metal dimers, such as EGaIn-Au and EGaIn-Ag dimers. **Figures S8** illustrates the absorption and scattering cross-section spectra for EGaIn-Au and EGaIn-Ag dimers, encompassing various dimer sizes ranging from 80 to 200 nm. When compared to single EGaIn nanospheres, which exhibit characteristics SPR in the UV range (ca. 200 nm), and single silver and gold spherical nanoparticles, which display SPR in the 400 and 500 nm range, the EGaIn-Au and EGaIn-Ag dimers exhibit prominent redshifted resonant peaks within the visible light range. The coupling-induced redshift is also strongly proportional to the size of dimers. Such strong coupling effect between soft EGaIn and rigid noble metal nanoparticles may open new applications in biosensing.

The aim of this study is to conduct a comprehensive investigation of the LSPR effects of EGaIn NPs by simulating the absorption and scattering power cross-section spectra for various shapes. A summary of all simulation results is presented in **Figure 7**, which illustrates a roadmap for the locations of major LSPR peaks of various EGaIn NPs, which have at least one dimension smaller than 100 nm. This map could be a quick guide for the researchers to determine the best shape of EGaIn NPs to achieve desired optical properties and therefore the optimal applications in various fields such as optical biosensors, medical bio-imaging, and drug delivery. For instance, for UV plasmonic applications, researchers may choose EGaIn nanospheres or nanoellipses for higher absorption efficiency. Similarly, for scattering contrast agents in the visible range, EGaIn nanorods seem a promising candidate due to balanced wavelength coverage and cross-section intensity. Overall, these major peaks in both scattering and absorption power cross-section spectra indicate that different shapes of EGaIn nanoparticles from 0D to 2D exhibit active SPR responses not only in the UV wavelength range, but also across visible to NIR wavelength regions. This is the first systematic study, demonstrating the unique optical activities of shape-controlled EGaIn NPs within the broad visible and NIR ranges. The SPR map can be used to guide the fabrication of various shapes of EGaIn nanoparticles to utilize their SPR advantages.

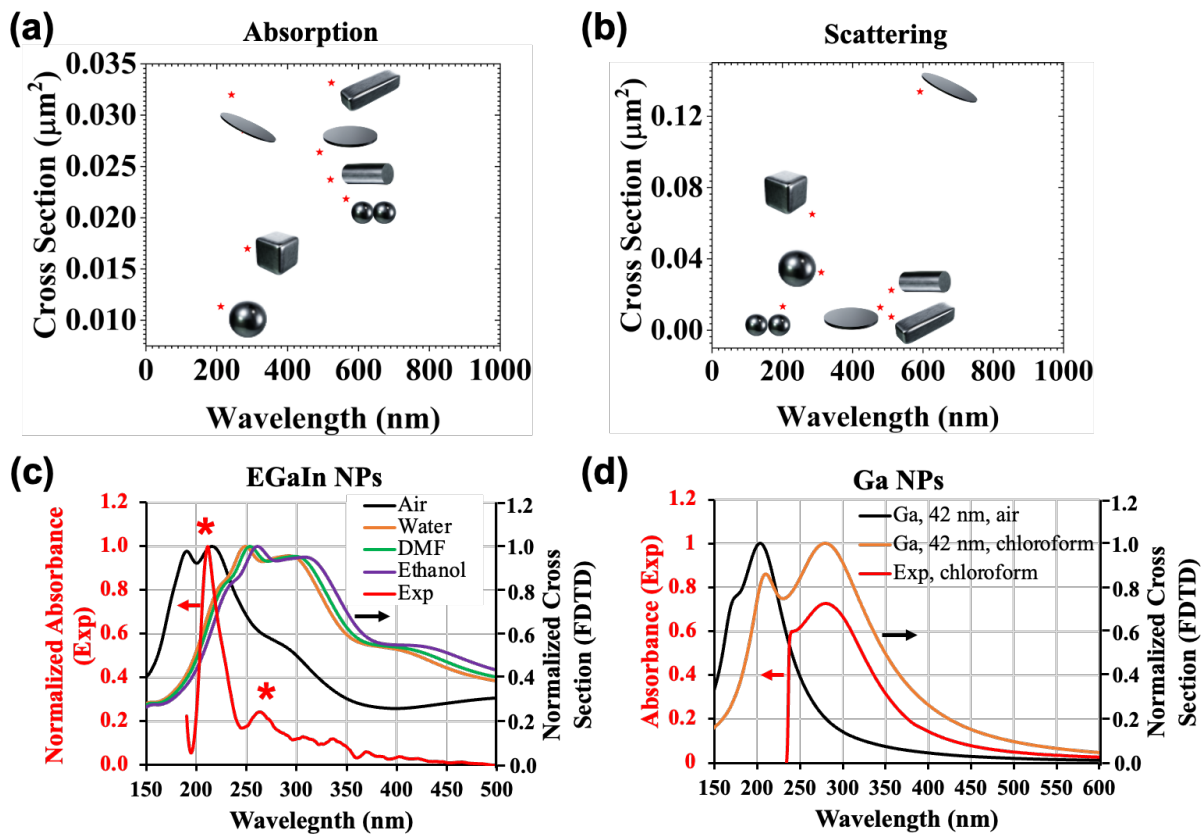

**Figure 7. Map of EGaIn nanoplasmonics and experimental validation.** Peak absorption (a) and scattering (b) wavelengths and corresponding cross-sections for different shapes that have been simulated by FDTD. (c) Overlap of experimental UV-vis absorption spectrum of EGaIn nanoparticles (185 nm by DLS) in ethanol with FDTD simulated absorption cross-section (160 nm) in different media (air, water, DMF, and ethanol). Red asterisks denote the major UV absorption peaks of synthesized EGaIn nanoparticles. (d) Overlap of experimental UV-vis absorption spectrum of Ga nanoparticles (42 nm by TEM) in chloroform with FDTD simulated absorption cross-section (42 nm) in different media (air and chloroform). The experimental UV-vis spectrum shows the UV cutoff effect of chloroform for wavelength shorter than ~245 nm.

We validated our FDTD simulation results by comparison with other simulation methods such as Mie theory (**Figure S3**) as well as real experimental measurements. The discrepancy between Mie theory and FDTD simulations in predicting the LSPR peak position for EGaIn nanospheres is reasonable and can be attributed to the differing assumptions and modeling capabilities of the two methods. Mie theory assumes ideal, homogeneous spherical particles in a uniform medium and does not account for retardation effects, surface oxidation (e.g., $Ga_2O_3$ shells), or substrate interactions. In contrast, FDTD simulations incorporate spatially resolved electromagnetic fields integrated and can model complex geometries, dielectric layering, and environmental influences with greater fidelity. Since we have experimental ellipsometry data for EGaIn in the 400–1000 nm range, future work will aim to apply the Drude–Lorentz[73] or Drude–Critical Points (DCP)[74] models to extrapolate the optical response into the ultraviolet (UV) region. Although these models are more complex and computationally demanding than the simple Drude model, they offer improved accuracy by capturing both intraband and interband electronic transitions—thereby enhancing the fidelity of FDTD simulations in the UV spectral range. For experimental validation, spherical EGaIn nanoparticles were fabricated by following a sonication method that we previously reported.[51] We used two different solvents, ethanol and dimethylformamide (DMF), for the synthesis. Based on the dynamic light scattering (DLS) analysis, the use of DMF as a dispersion solvent resulted in a smaller diameter size of nanoparticles compared to ethanol (**Figure S9a**), and also weaker UV absorption with higher measurement noise (**Figure S9b**). The UV-vis spectrum of EGaIn nanoparticles in ethanol was further overlapped with the FDTD simulation results (**Figure 7c**). Both experimental data and computational analysis showed two clear absorption peaks in the UV range (**Figure 7c**). The simulation results also showed a red-shift of UV absorption peaks of EGaIn as the refractive index of the medium increased (air – 1, water – 1.4, DMF – 1.43, and ethanol –1.478). The experimental UV absorption peaks of synthesized EGaIn nanoparticles (red asterisks, **Figure 7c**) are slightly blue-shifted compared to the ethanol simulation results (purple curve, **Figure 7c**), and fall in between the FDTD simulation results obtained using air and water as media. This could be due to the small particle size difference between experimental (185 nm, by DLS) and theoretical (160 nm) results. Additionally, we synthesized two different sizes of Ga nanoparticles: 42 nm in chloroform and 27 nm in hexane (**Figure S10**). Hexane was used for smaller Ga nanoparticles to minimize UV absorption below 245 nm. For Ga nanoparticles, much better matching between the experimental measurement and FDTD predication was observed for both size particles (**Figure 7d** and **Figure S11**). Current validation is limited to spherical LM nanoparticles. When the continuous development of advanced fabrication methods for LMs, the SPR features of more complicate shaped LMs could be validated experimentally in future.

The simulated red-shift in the EGaIn SPR with increasing nanoparticle size (**Figure 1c**) and surrounding refractive index (**Figure 7c**) can be theoretically explained by classical electrodynamics. As the diameter of EGaIn nanoparticles increases, both polarizability and radiative damping become more significant. These effects reduce the restoring force on the oscillating conduction electrons, resulting in a lower resonance frequency and, consequently, a red-shift in the SPR peak. This phenomenon is well-described by Mie theory[62] and the Drude[75] model, which predict size-dependent plasmonic behavior consistent with our FDTD simulations. Moreover, changes in the surrounding dielectric medium also influence the plasmonic response. According to the Fröhlich condition[76], an increase in the refractive index of the surrounding medium shifts the dipolar surface plasmon resonance toward lower photon energies, resulting in a red shift of the absorption peak. This sensitivity to the dielectric environment further confirms the suitability of EGaIn nanoparticles for refractive index-based sensing applications, including bio- and chemical detection platforms.

To date, various shapes of LM NPs have been successfully fabricated using different techniques. For instance, EGaIn spherical NPs were synthesized through sonication[52]. EGaIn nanorods, measuring 210 nm in diameter and 850 nm in length, were created using ultrasound-assisted physical dispersion[67]. Arrays of EGaIn nanowires were formed by pulling an EGaIn droplet on a flat substrate at room temperature, resulting in an hourglass-shaped structure that evolved into nanowires through a breakdown process[68]. EGaIn nanodroplets were synthesized through polymeric ligand encapsulation, involving the covalent bonding of various polymers[77]. Additionally, a recent nanofabrication technique took advantage of intermetallic wetting between Au and EGaIn LM on a Si/SiO$_2$ substrate, yielding LM nanostructures of different 2D geometries such as square, circle, and hexagonal disk shapes[78]. The fabricated LM nanostructures exhibited smooth thin-film surfaces, sharp edges, and corners, with lateral feature sizes down to ~500 nm and the smallest spacing between array elements demonstrated to date (~250 nm). The LM nanostructures showed strong resonant optical responses in the broad UV-vis-NIR region, as indicated by the FDTD results (**Figure 7**). Such rich SPR properties will also promote the development of new fabrication technologies to generate shapes that have not been fabricated before. In particular, technology that can transform the shape of EGaIn nanoparticles is even more interesting, as that achieves the ultimate goal of transformable plasmonics. Integrating the new fabrication techniques and experimental measurements with FDTD simulation to confirm the SPR properties of EGaIn NPs could be one of future directions. The collaborative results will further convince LM's emerging applications in biosensing and many others[79–83].

- **Conclusions**

To summarize, this study provides a comprehensive exploration of shape and size-dependent LSPR properties of EGaIn NPs in the broad UV-vis-NIR spectral range. Through FDTD simulations, we have investigated the LSPR effects exhibited by EGaIn nanoparticles across various shapes, including 0D, 1D, 2D, and plasmonic coupling formed by nanosphere dimers. The size-dependent SPR effects observed in the UV, visible, and NIR light ranges highlight the significance of tailored nanoparticle geometry in achieving desired optical and plasmonic characteristics. The presented guide map of major peaks and absorption and scattering power cross-sections for different shapes of EGaIn nanoparticles serves as a valuable resource for researchers seeking optimal configurations for specific applications, including biosensing, medical bio-imaging, drug delivery, and beyond. This research not only contributes to the fundamental understanding of EGaIn NP behavior but also underscores the potential for shape-transformable gallium-based nanoparticles. As the field of "liquid plasmonics" continues to evolve, the findings presented here pave the way for further advancements in nanofabrication techniques and the development of novel applications in diverse scientific and technological domains.

- **Supporting information**

Supporting Information is available from the Wiley Online Library or from the author.

- **Acknowledgement**

The authors sincerely thank the funding support from the National Science Foundation (Award # 1944167).

- **Conflicts of interest**

The authors declare no competing financial interests.

- **Author contribution**

Project conception and visualization were performed by SJ, MDD, and QW. FDTD computational simulation and Mie theory computational study were conducted by SJ. Data analysis was performed by SJ. MZ assisted with the introduction section of the manuscript and performed the UV-vis measurements of EGaIn nanoparticles. MJD synthesized the spherical EGaIn nanoparticles and conducted the DLS measurements. FS and MY provided the Ga nanoparticles and performed TEM and UV-vis analysis. AV designed 3D schematic shapes of EGaIn with Blender. SJ, MDD, and QW wrote the manuscript. All authors contributed to the revision of the manuscript.

- **Data availability**

The data that support the findings of the study are available in the supporting material or from the corresponding authors upon request.

- **References**

1  X. Han, K. Liu and C. Sun, *Materials*, 2019, **12**, 1411.
2  S. Maier, *Plasmonics: Fundamentals and Applications*, 2007.
3  K. L. Kelly, E. Coronado, L. L. Zhao and G. C. Schatz, *J. Phys. Chem. B*, 2003, **107**, 668–677.
4  L. Guo, S. Xu, X. Ma, B. Qiu, Z. Lin and G. Chen, *Sci. Rep.*, 2016, **6**, 32755.
5  H. Im, H. Shao, Y. I. Park, V. M. Peterson, C. M. Castro, R. Weissleder and H. Lee, *Nat. Biotechnol.*, 2014, **32**, 490–495.
6  S. Wang, E. S. Forzani and N. Tao, *Anal. Chem.*, 2007, **79**, 4427–4432.
7  Z. Zou, Y. Chen, S. Yuan, N. Luo, J. Li and Y. He, *Adv. Funct. Mater.*, 2023, **33**, 2213312.
8  K. Kurzątkowska, T. Santiago and M. Hepel, *Biosens. Bioelectron.*, 2017, **91**, 780–787.
9  Surface Modification of Gallium-Based Liquid Metals: Mechanisms and Applications in Biomedical Sensors and Soft Actuators - Kwon - 2021 - Advanced Intelligent Systems - Wiley Online Library, https://onlinelibrary.wiley.com/doi/full/10.1002/aisy.202000159, (accessed November 9, 2023).
10 R. Bakhtiar, *J. Chem. Educ.*, 2013, **90**, 203–209.
11 D. R. Shankaran, K. V. Gobi and N. Miura, *Sens. Actuators B Chem.*, 2007, **121**, 158–177.
12 M. I. Stockman, *Opt Express*, DOI:10.1364/OE.19.022029.
13 M. W. Knight, *ACS Nano*, DOI:10.1021/nn405495q.
14 N. Liu, M. L. Tang, M. Hentschel, H. Giessen and A. P. Alivisatos, *Nat Mater*, DOI:10.1038/nmat3029.
15 D. Y. Fedyanin, D. I. Yakubovsky, R. V. Kirtaev and V. S. Volkov, *Nano Lett*, DOI:10.1021/acs.nanolett.5b03942.
16 Y. Yang, *ACS Photonics*, DOI:10.1021/ph500042v.
17 J. M. McMahon, G. C. Schatz and S. K. Gray, *Phys Chem Chem Phys*, DOI:10.1039/C3CP43856B.
18 P. C. Wu, *Appl Phys Lett*, DOI:10.1063/1.2712508.
19 T. Zhang, *Opt Mater Express*, DOI:10.1364/OME.7.002880.
20 G. V. Naik, V. M. Shalaev and A. Boltasseva, *Adv Mater*, DOI:10.1002/adma.201205076.
21 J. M. Sanz, *J Phys Chem C*, DOI:10.1021/jp405773p.
22 K. Appusamy, S. Blair, A. Nahata and S. Guruswamy, *Mater Sci Eng B*, DOI:10.1016/j.mseb.2013.11.009.
23 J. Toudert and R. Serna, *Opt Mater Express*, DOI:10.1364/OME.6.002434.
24 A. Cuadrado, J. Toudert and R. Serna, *IEEE Photonics J*, DOI:10.1109/JPHOT.2016.2574777.
25 D. Morales, N. A. Stoute, Z. Yu, D. E. Aspnes and M. D. Dickey, *Appl Phys Lett*, DOI:10.1063/1.4961910.
26 I. S. Maksymov, *Rev Phys*, DOI:10.1016/j.revip.2016.03.002.
27 Recent advances in plasmonic nanostructures for sensing: a review, https://www.spiedigitallibrary.org/journals/optical-engineering/volume-54/issue-10/100902/Recent-advances-in-plasmonic-nanostructures-for-sensing-a-review/10.1117/1.OE.54.10.100902.full, (accessed June 5, 2025).
28 Y. Lu, *Nat Commun*.
29 A. G. Marín, *Nanoscale*, DOI:10.1039/C6NR00926C.
30 P. Strobbia and B. M. Cullum, *SPIE Opt Eng*, DOI:10.1117/1.OE.54.10.100902.
31 H. Fatakdawala, *J Cardiovasc Trans Res*, DOI:10.1007/s12265-015-9627-3.
32 C. V. Bourantas, *Eur Heart J*, DOI:10.1093/eurheartj/ehw097.

33  B. Wang, *IEEE J Quantum Electron*, DOI:10.1109/JSTQE.2009.2037023.
34  M. W. Knight, *ACS Nano*, DOI:10.1021/nn5072254.
35  Y. Lin, Y. Liu, J. Genzer and M. D. Dickey, *Chem Sci*, DOI:10.1039/C7SC00057J.
36  Z. Huang, M. Guan, Z. Bao, F. Dong, X. Cui and G. Liu, *Small*, 2024, **20**, 2306652.
37  W. Sun, J. Nan, Y. Che, H. Shan, Y. Sun, W. Xu, S. Zhu, J. Zhang and B. Yang, *Biosens. Bioelectron.*, 2024, **261**, 116469.
38  Stretchable and Soft Electronics using Liquid Metals - Dickey - 2017 - Advanced Materials - Wiley Online Library, https://onlinelibrary.wiley.com/doi/abs/10.1002/adma.201606425, (accessed November 9, 2023).
39  S.-Y. Tang, C. Tabor, K. Kalantar-Zadeh and M. D. Dickey, *Annu. Rev. Mater. Res.*, 2021, **51**, 381–408.
40  T. Daeneke, K. Khoshmanesh, N. Mahmood, I. A. de Castro, D. Esrafilzadeh, S. J. Barrow, M. D. Dickey and K. Kalantar-zadeh, *Chem. Soc. Rev.*, 2018, **47**, 4073–4111.
41  Liquid Metal–Based Soft Microfluidics - Zhu - 2020 - Small - Wiley Online Library, https://onlinelibrary.wiley.com/doi/abs/10.1002/smll.201903841, (accessed November 9, 2023).
42  M. D. Dickey, *ACS Appl. Mater. Interfaces*, 2014, **6**, 18369–18379.
43  K. Kalantar-Zadeh, J. Tang, T. Daeneke, A. P. O'Mullane, L. A. Stewart, J. Liu, C. Majidi, R. S. Ruoff, P. S. Weiss and M. D. Dickey, *ACS Nano*, 2019, **13**, 7388–7395.
44  A Short History of Fusible Metals and Alloys – Towards Room Temperature Liquid Metals - Handschuh-Wang - 2022 - European Journal of Inorganic Chemistry - Wiley Online Library, https://chemistry-europe.onlinelibrary.wiley.com/doi/abs/10.1002/ejic.202200313, (accessed November 9, 2023).
45  Y. Lin, J. Genzer and M. D. Dickey, *Adv. Sci.*, 2020, **7**, 2000192.
46  W. Babatain, M. S. Kim and M. M. Hussain, *Adv. Funct. Mater.*, **n/a**, 2308116.
47  S. Jamalzadegan, S. Kim, N. Mohammad, H. Koduri, Z. Hetzler, G. Lee, M. D. Dickey and Q. Wei, *Adv. Funct. Mater.*, **n/a**, 2308173.
48  Band structure and optical properties of gallium - IOPscience, https://iopscience.iop.org/article/10.1088/0305-4608/4/11/032/meta, (accessed November 9, 2023).
49  A. Bhardwaj and S. S. Verma, *Mater. Today Commun.*, 2023, **37**, 107135.
50  A. Bhardwaj, P. Bhatia and S. S. Verma, *Opt. Quantum Electron.*, 2022, **55**, 40.
51  A. Bhardwaj and S. S. Verma, *J. Quant. Spectrosc. Radiat. Transf.*, 2022, **281**, 108109.
52  P. Reineck, Y. Lin, B. C. Gibson, M. D. Dickey, A. D. Greentree and I. S. Maksymov, *Sci. Rep.*, 2019, **9**, 5345.
53  J. Ma, F. Krisnadi, M. H. Vong, M. Kong, O. M. Awartani and M. D. Dickey, *Adv. Mater.*, **n/a**, 2205196.
54  P. Q. Liu, X. Miao and S. Datta, *Opt. Mater. Express*, 2023, **13**, 699–727.
55  Z. Yu, *Phys Rev Lett*, DOI:10.1103/PhysRevLett.121.024302.
56  T. R. Lear, *Extrem Mech Lett*, DOI:10.1016/j.eml.2017.02.009.
57  S. Connor, DOI:10.1109/ISEMC.2008.4652175.
58  S. van der Veeke, bachelor, Faculty of Science and Engineering, 2014.
59  X. Wu, H. Fang, X. Ma and S. Yan, *Adv. Opt. Mater.*, 2023, **11**, 2301180.
60  D. Morales, N. A. Stoute, Z. Yu, D. E. Aspnes and M. D. Dickey, *Appl. Phys. Lett.*, 2016, **109**, 091905.
61  D. Debnath and S. K. Ghosh, *ACS Appl. Nano Mater.*, 2022, **5**, 1621–1634.

62 Q. Fu and W. Sun, *Appl. Opt.*, 2001, **40**, 1354–1361.
63 I. D. Joshipura, C. K. Nguyen, C. Quinn, J. Yang, D. H. Morales, E. Santiso, T. Daeneke, V. K. Truong and M. D. Dickey, *iScience*, 2023, **26**, 106493.
64 L. Liu, Y. Wu, N. Yin, H. Zhang and H. Ma, *J. Quant. Spectrosc. Radiat. Transf.*, 2020, **240**, 106682.
65 M. A. Mahmoud, M. Chamanzar, A. Adibi and M. A. El-Sayed, *J. Am. Chem. Soc.*, 2012, **134**, 6434–6442.
66 J. Liu and Z. Li, *Photonics*, 2022, **9**, 53.
67 Z. Li, H. Zhang, D. Wang, C. Gao, M. Sun, Z. Wu and Q. He, *Angew. Chem. Int. Ed.*, 2020, **59**, 19884–19888.
68 T. Ikuno and Z. Somei, *Molecules*, 2021, **26**, 4616.
69 E.-A. You, W. Zhou, J. Y. Suh, M. D. Huntington and T. W. Odom, *ACS Nano*, 2012, **6**, 1786–1794.
70 C.-C. Huang, C.-S. Yeh and C.-J. Ho, *J. Phys. Chem. B*, 2004, **108**, 4940–4945.
71 J. H. Yoon, F. Selbach, L. Schumacher, J. Jose and S. Schlücker, *ACS Photonics*, 2019, **6**, 642–648.
72 E. Hao and G. C. Schatz, *J. Chem. Phys.*, 2003, **120**, 357–366.
73 H. S. Sehmi, W. Langbein and E. A. Muljarov, *Phys. Rev. B*, 2017, **95**, 115444.
74 K. P. Prokopidis and D. C. Zografopoulos, *J. Light. Technol.*, 2013, **31**, 2467–2476.
75 E. Silaeva, L. Saddier and J.-P. Colombier, *Appl. Sci.*, 2021, **11**, 9902.
76 X. Fan, W. Zheng and D. J. Singh, *Light Sci. Appl.*, 2014, **3**, e179–e179.
77 J. Yan, M. H. Malakooti, Z. Lu, Z. Wang, N. Kazem, C. Pan, M. R. Bockstaller, C. Majidi and K. Matyjaszewski, *Nat. Nanotechnol.*, 2019, **14**, 684–690.
78 M. A. K. Khan, Y. Zhao, S. Datta, P. Paul, S. Vasini, T. Thundat and P. Q. Liu, *Small*, **n/a**, 2403722.
79 O. Hossain, Y. Wang, M. Li, B. Mativenga, S. Jamalzadegan, N. Mohammad, A. Velayati, A. D. Poonam and Q. Wei, *Biosens. Bioelectron.*, 2025, **278**, 117341.
80 Z. Hetzler, Y. Wang, D. Krafft, S. Jamalzadegan, L. Overton, M. W. Kudenov, F. S. Ligler and Q. Wei, *Front. Chem.*
81 G. Lee, O. Hossain, S. Jamalzadegan, Y. Liu, H. Wang, A. C. Saville, T. Shymanovich, R. Paul, D. Rotenberg, A. E. Whitfield, J. B. Ristaino, Y. Zhu and Q. Wei, *Sci. Adv.*, 2023, **9**, eade2232.
82 S. Im, E. Frey, D. H. Kim, S.-Y. Heo, Y. M. Song, M. H. Vong, S. Jamalzadegan, Q. Wei, A. A. Gregg, O. Khatib, W. J. Padilla, J. Genzer and M. D. Dickey, *Adv. Funct. Mater.*, **n/a**, 2422453.
83 S. Jamalzadegan, J. Xu, Y. Shen, B. Mativenga, M. Li, M. Zare, A. Penumudy, Z. Hetzler, Y. Zhu and Q. Wei, *Chem Bio Eng.*, DOI:10.1021/cbe.5c00027.

# Supplementary Information

## Shape and size-dependent surface plasmonic resonances of liquid metal alloy (EGaIn) nanoparticles

*Sina Jamalzadegan[1], Mohammadreza Zare [1], Micah J. Dickens[1], Florian Schenk [2], Alireza Velayati[1], Maksym Yarema[2], Michael D. Dickey[1]\*, and Qingshan Wei[1]\**

[1] Department of Chemical and Biomolecular Engineering, North Carolina State University, Raleigh, NC 27695, USA

[2] Department of Information Technology and Electrical Engineering, ETH Zürich, 8092 Zürich, Switzerland

\* Co-corresponding: qwei3@ncsu.edu and mddickey@ncsu.edu

**MATERIALS.** Gallium tris(dimethylamido) dimer ($Ga_2(DMA)_6$, 98%) was purchased from American Elements and used as received. Dioctylamine (DOA, 98%), didodecylamine (DDA, >97%), 1-octadecene (1-ODE, technical, 90%), oleic acid (90 %), chloroform (≥ 99.8%) were bought from Sigma-Aldrich. DOA, DDA and 1-ODE were dried and degassed at 110 °C under dynamic vacuum and transferred air-free into a nitrogen-filled glovebox. Hexane and ethanol were purchased from VWR. All synthesis steps were performed using conventional Schlenk and glove box techniques under nitrogen.

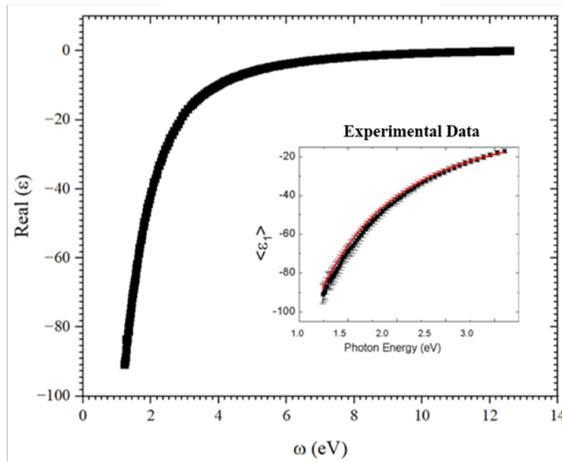
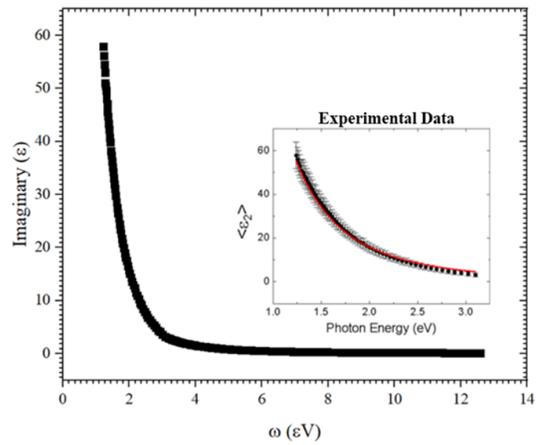

**Figure S1.** Extrapolated and experimental[1] values (insets) for (a) the real part and (b) the imaginary part of the permittivity of EGaIn in the 100–1000 nm wavelength range.

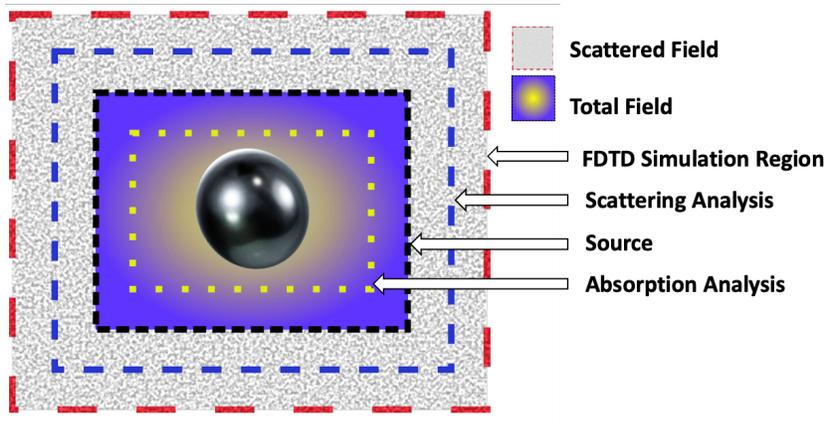
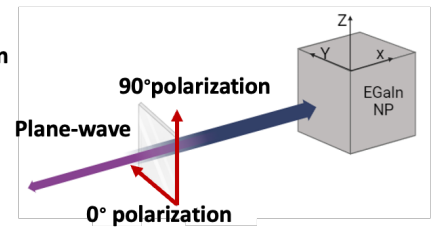

**Figure S2.** (a) Schematic of FDTD simulation Setup, (b) Schematic of the incidence light source and polarization angle.

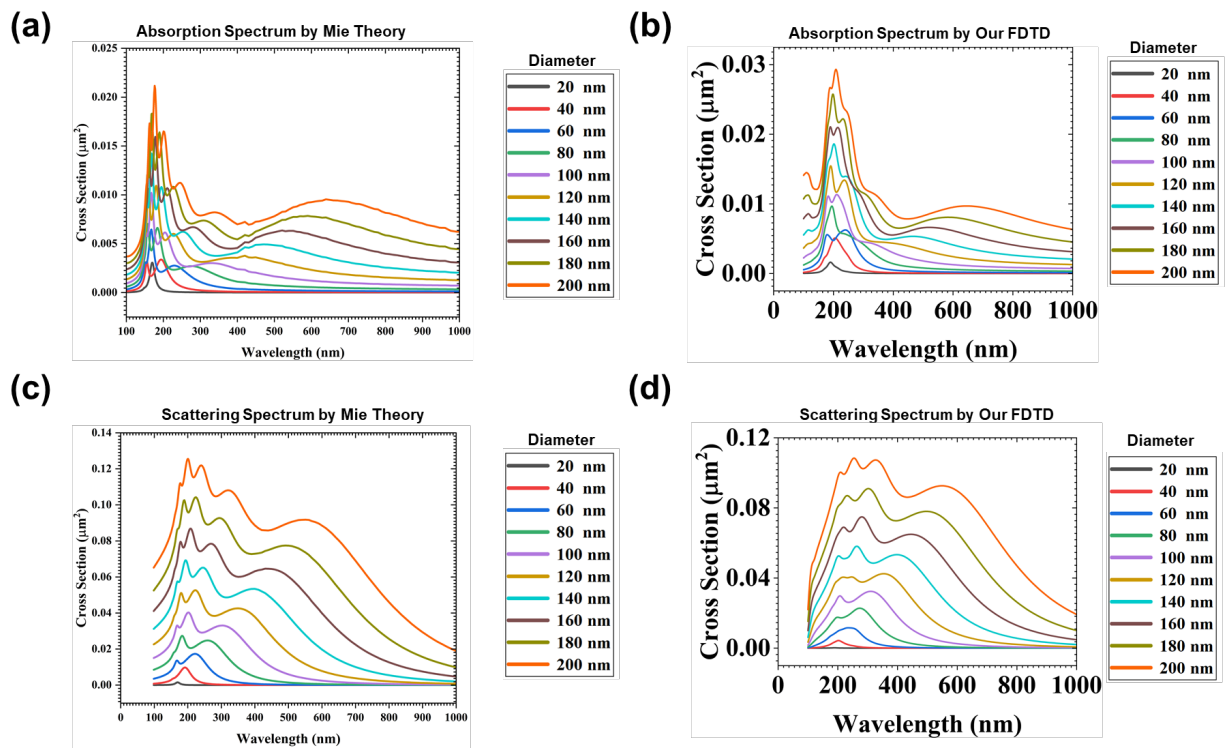

**Figure S3.** Absorption spectra of various diameters of single EGaIn spherical nanoparticles by applying (a) Mie theory and (b) FDTD simulation. Scattering spectra of various diameters of single EGaIn spherical nanoparticles by applying (c) Mie theory and d) FDTD simulation.

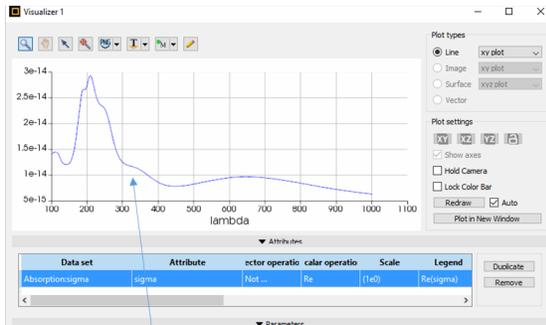
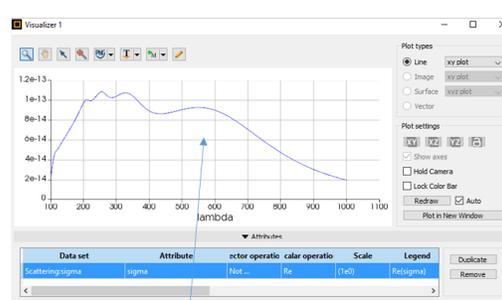
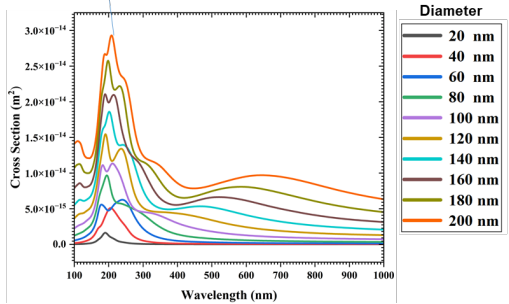
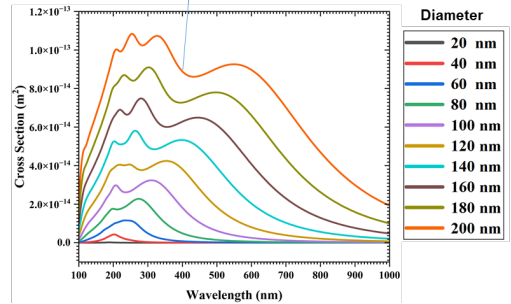

**Figure S4.** The absorption and scattering spectra of a single spherical EGaIn nanoparticle with a 200 nm diameter and a 6 nm oxide layer show no changes compared to a single spherical EGaIn nanoparticle without an oxide layer.

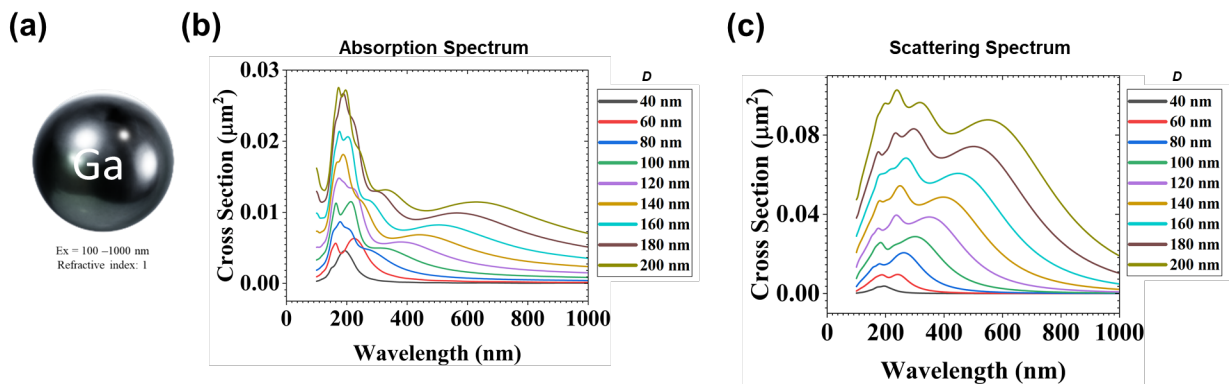

**Figure S5.** (a) 3D schematic of a single spherical Ga nanoparticle. FDTD simulation results of (b) absorption and (c) scattering spectra of various diameters of single Ga spherical nanoparticles. Ga nanosphere is excited by a plane-wave source of 100-1000 nm wavelength in the x-direction with a polarized angle of 90 degree.

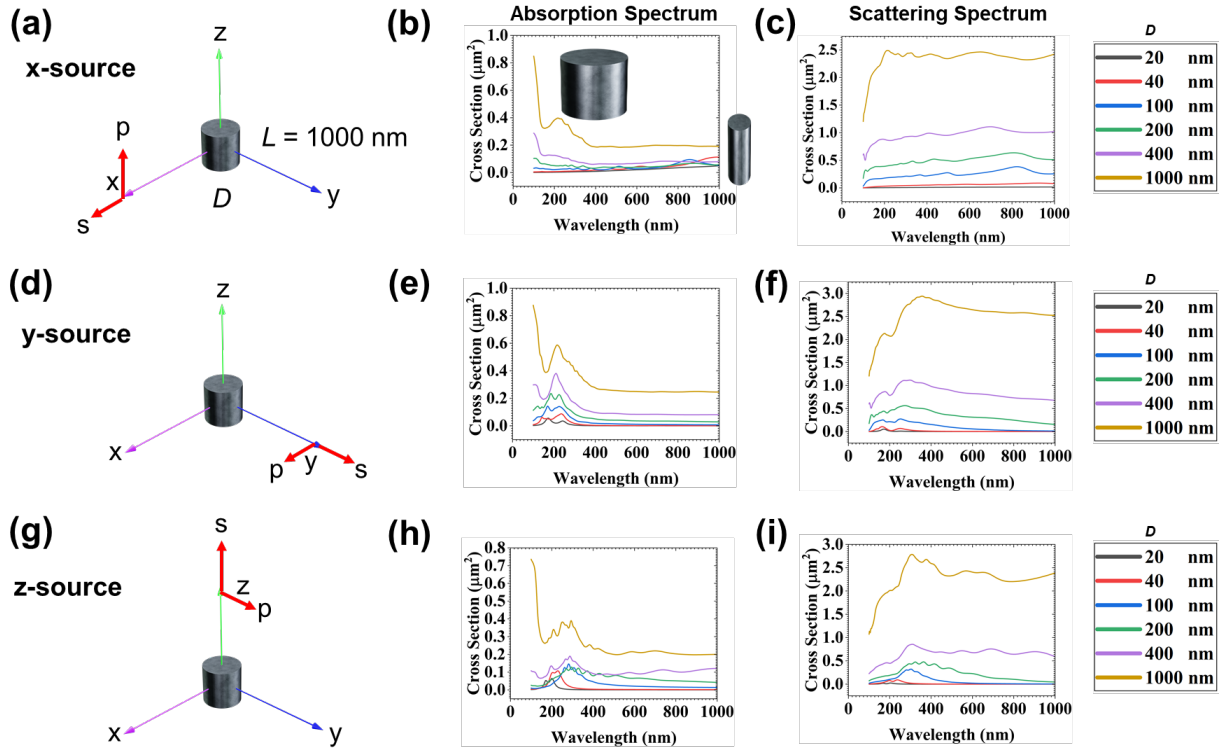

**Figure S6. Plasmonic resonance behavior of single EGaIn nanorod with tunable diameters.** (a), (d), and (g) 3D schematic of single EGaIn nanorod excited by a plane-wave light source (s) in x-, y-, and z-direction with 90 degree polarization (p), respectively. (b), (e), (h) Absorption and (c), (f), (i) scattering cross section spectra for various diameters with a constant length of 1000 nm.

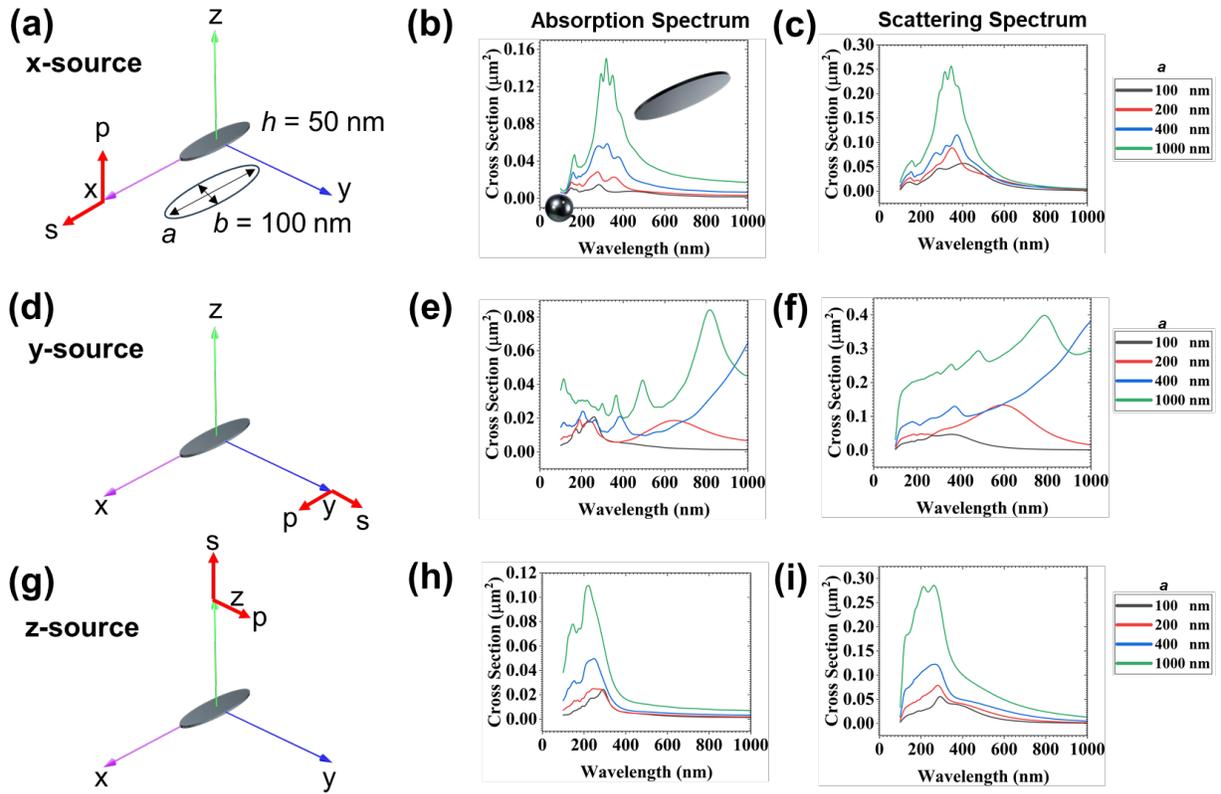

**Figure S7. Plasmonic resonance behavior of single EGaIn nanoellipse.** (a), (d), and (g) 3D schematic of single EGaIn nanoellipse excited by a plane-wave light source (s) in x-, y-, and z- directions with 90 degree polarization (p), respectively. (b), (e), (h) Absorption and (c), (f), and (i) scattering cross section spectra for nanoellipses with a constant thickness of 50 nm and minor axis of 100 nm.

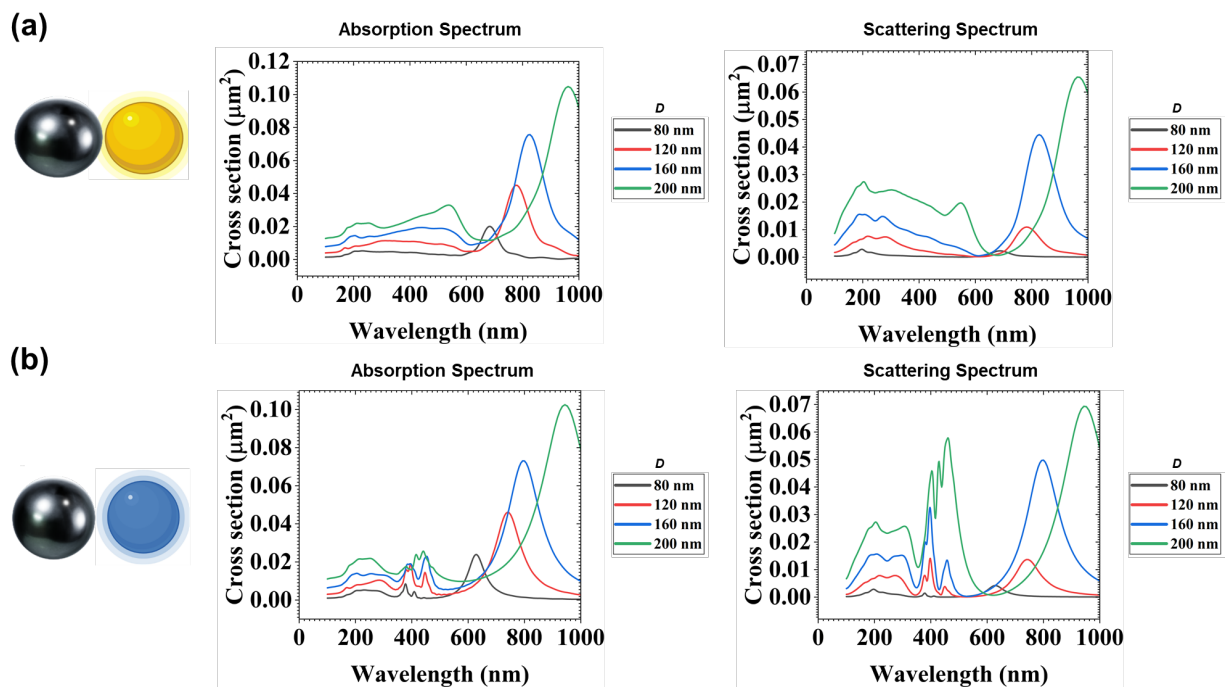

**Figure S8.** (a) Absorption and (b) scattering spectrum of various radius of dimer EGaIn-Au spherical nanoparticles and also (c) Absorption and (d) scattering spectrum various radius of dimer EGaIn-Ag spherical applying FDTD simulation by using plane-wave source with 100-1000 nm wavelength in x-direction with polarized angle 90 degree.

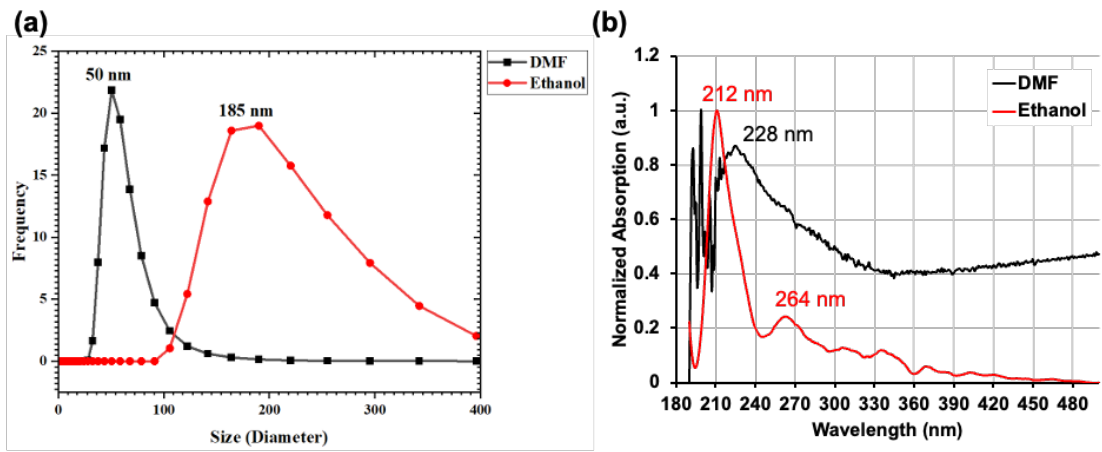

**Figure S9. Characterization of experimentally synthesized EGaIn spherical nanoparticles.** (a) Dynamic light scattering (DLS) measurements of the synthesized spherical EGaIn nanoparticles in dispersed in ethanol (red circle) and dimethylformamide (DMF, black square) solvents. (b) UV-vis absorption spectra of synthesized EGaIn nanoparticles in ethanol (red) and DMF (black), respectively.

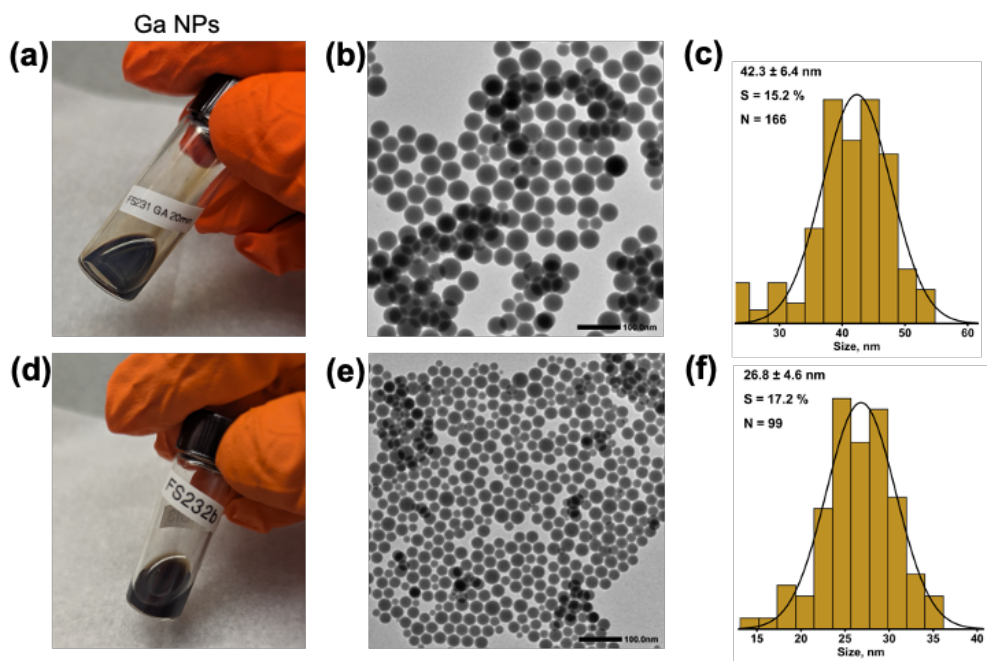

**Figure S10.** Characterization of two Ga NPs: 42 nm Ga NPs dispersed in chloroform and 27 nm Ga NPs dispersed in hexane, respectively. (a,d) Photographs of synthesized samples. (b,e) TEM images. (c,f) Size analysis based on TEM images.

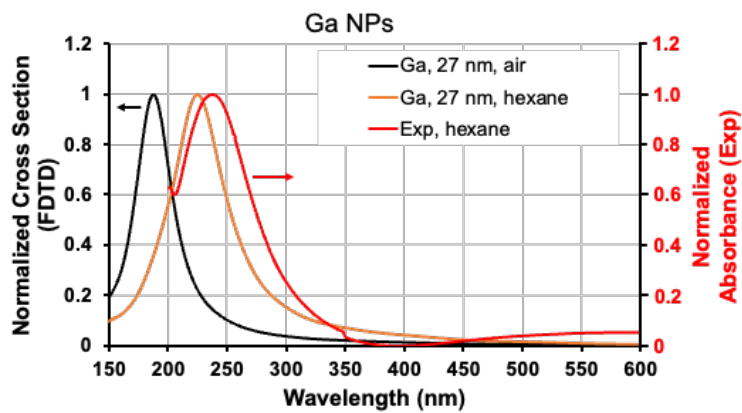

**Figure S11.** Overlap of experimental UV-vis absorption spectrum of Ga nanoparticles (27 nm by TEM) in hexane with FDTD simulated absorption cross-section (27 nm) in different media (air and hexane).

# REFERENCES

1) Morales, D.; Stoute, N. A.; Yu, Z.; Aspnes, D. E.; Dickey, M. D. Liquid Gallium and the Eutectic Gallium Indium (EGaIn) Alloy: Dielectric Functions from 1.24 to 3.1 eV by Electrochemical Reduction of Surface Oxides. *Appl Phys Lett* **2016**, *109*. https://doi.org/10.1063/1.4961910.